\documentclass[a4paper, 11pt, fleqn]{article}
\usepackage{mystyle}
\usepackage{cancel}

\providecommand{\given}{\ensuremath{\hspace{-0.1em}\mid\hspace{-0.1em}}}

\providecommand{\trans}{\ensuremath{^\mathsf{T}}}
\providecommand{\inv}{\ensuremath{^{-1}}}

\providecommand{\parallax}{\ensuremath{\varpi}}

\providecommand{\propm}{\ensuremath{\mu}}
\providecommand{\pmra}{\ensuremath{\mu_{\alpha}}}
\providecommand{\pmdec}{\ensuremath{\mu_{\delta}}}
\providecommand{\sigparallax}{\ensuremath{\sigma_{\varpi}}}
\providecommand{\sigpmra}{\ensuremath{\sigma_{\pmra}}}
\providecommand{\sigpmdec}{\ensuremath{\sigma_{\pmdec}}}
\providecommand{\corparallaxpmra}{\ensuremath{\rho_{\parallax\pmra}}}
\providecommand{\corparallaxpmdec}{\ensuremath{\rho_{\parallax\pmdec}}}
\providecommand{\corpmrapmdec}{\ensuremath{\rho_{\pmra\pmdec}}}
\providecommand{\cov}{\ensuremath{\Sigma}}

\providecommand{\hp}{\ensuremath{p}} 
\providecommand{\dist}{\ensuremath{r}}

\providecommand{\vra}{\ensuremath{v_{\alpha}}}
\providecommand{\vdec}{\ensuremath{v_{\delta}}}

\providecommand{\kms}{\ensuremath{\mathrm{km}\,\mathrm{s}^{-1}}}

\providecommand{\kpc}{\ensuremath{\mathrm{kpc}}}
\providecommand{\mas}{\ensuremath{\mathrm{mas}}}
\providecommand{\maspyr}{\ensuremath{\mathrm{mas\,\mathrm{yr}^{-1}}}}

\begin{document}
\thispagestyle{empty}

{\bfseries\Large
Estimating distances from parallaxes. VII.\\[5pt]
Computing the three-dimensional kinegeometric posterior with\\[5pt]
one-dimensional sampling
}

\vspace{5mm}
{\large Coryn A.L.\ Bailer-Jones\\
 Max Planck Institute for Astronomy, Heidelberg\\
 \url{https://www.mpia.de/homes/bailer-jones}\\
12 November 2026

%%%%%%%%%%%%%%%%%%%%%%%%%%%%%%%%%%%%%%%%%%%%%%%%
 \section*{Abstract}

When parallaxes are of poor quality, proper motions can improve stellar distance estimates using known constraints on stars' physical velocities. Formally this may be achieved via the kinegeometric posterior, a three-dimensional (3D) posterior in distance and velocity for given measured parallax and 2D proper motion. This posterior is the product of a covariant Gaussian likelihood in the parallax and proper motion with a distance and velocity prior that can be constructed from a Galaxy model. In earlier work, I showed how to infer this posterior for stars in Gaia DR3 by explicitly sampling it with a Markov Chain Monte Carlo method. Here I show that this posterior is equivalent to a 1D posterior in distance combined with a sum of 2D Gaussian distributions in velocity. This 1D+2D representation is both faster and more accurate to sample, thereby making the computation of the kinegeometric posterior for a large set of sources much easier.

%%%%%%%%%%%%%%%%%%%%%%%%%%%%%%%%%%%%%%%%%%%%%%%%
\section{Introduction}

Distances to individual stars can be inferred from parallaxes via a Bayesian method that yields meaningful point and uncertainty estimates even with very noisy or negative parallaxes (Bailer-Jones 2015; see also Luri et al.\ 2018 and Queiroz et al.\ 2018). By combining the parallax likelihood with a suitable prior, a distance posterior can be computed with a width that reflects a combination of the uncertainty in the parallax and the width of the prior. I used this method to construct distance catalogues based on the parallaxes published in the second (Bailer-Jones et al.\ 2018) and third (Bailer-Jones et al.\ 2021, hereafter paper V) Gaia data releases (Gaia Collaboration 2018, 2021). The distance estimates that use only the parallax measurement are called {\em geometric} distances.

Although Gaia parallaxes are generally very accurate, they barely constrain the distance to stars more than several kiloparsecs away, with the result that the distance posteriors are dominated by the prior. However, even at these large distances, the stars' proper motions are often still comparatively accurate, and when combined with prior information on the velocities, they provide independent constraints on distance. In Bailer-Jones (2023, hereafter paper~VI), I showed how to infer stellar distances and transverse velocities from the parallax and proper motions and demonstrated this using data from Gaia DR3.\footnote{Appendix A of paper~VI illustrates how the proper motion provides information on the distance even without a parallax. It also shows that the extent to which even very precise proper motions can constrain distance is limited by the width of the velocity prior (see also section 5 of that paper).}
As these distance estimates depend both on geometry (parallax) and kinematics (proper motions), they are called {\em kinegeometric} distances.
In paper~VI I directly sampled the joint three-dimensional (3D) distance--velocity posterior with a Markov Chain Monte Carlo (MCMC) method. While doable, this approach is relatively slow and may also be unreliable if insufficient samples are used or if the covariances in the data are very large.

Here I show that, because velocity is linearly proportional to proper motion at a given distance, the 3D sampling can be reduced without approximation to a 1D sampling of the distance posterior followed by single samplings of multiple 2D conditional velocity posteriors.
By further adopting a Gaussian prior on the velocities at a given distance (as also done in paper~VI), these 2D distributions become Gaussian, so the velocity posterior becomes a sum of Gaussians. This 1D+2D sampling method is equivalent to sampling the full 3D posterior, but faster and more accurate.

%%%%%%%%%%%%%%%%%%%%%%%%%%%%%%%%%%%%%%%%%%%%%%%%
\section{A note on notation}

The parallax is denoted \parallax\ and the distance \dist.
Both the proper motion and velocity are 2D vectors, whereby the two proper motion components are $\propm = (\pmra,\pmdec)$ and the two corresponding velocity components are $v = (\vra, \vdec)$.
I write the proper motion in right ascension as \pmra\ rather than $\mu_{\alpha*}$, but it has angular units per unit time, i.e.\ it is adjusted for declination.
$\mathcal{N}_x(m_x, \cov_x)$ denotes a (generally multivariate) Gaussian in $x$ with mean $m_x$ and covariance $\cov_x$. $P$ denotes a probability density function; an asterisk $P^*$ indicates it is not normalized.

%%%%%%%%%%%%%%%%%%%%%%%%%%%%%%%%%%%%%%%%%%%%%%%%
\section{Derivation}
 
A noise-free proper motion \pmra\ (angular velocity) is related to the distance \dist\ and transverse velocity \vra\ via
\begin{alignat}{2}
  \frac{\vra}{\kms} \,=&\, k \ \frac{\dist}{\kpc} \frac{\pmra}{\maspyr} \hspace*{0.5em}\textrm{with}\hspace*{0.5em}k=4.740471 
\label{eqn:transvel}
\end{alignat}
and likewise for the other proper motion and velocity component.

The unnormalized 3D posterior probability density function (PDF) over distance and velocity is the product of the likelihood and prior.
The joint likelihood in parallax and proper motion for Gaia is well represented by a 3D Gaussian (Lindegren et al.\ 2018) with mean and covariance
\begin{subequations}\label{eqn:like_3d}
\begin{alignat}{2}
  m_3\,&=\, \begin{bmatrix}m_\parallax, m_\propm\end{bmatrix}\trans \,=\, \begin{bmatrix}\frac{1}{\dist}, \frac{\vra}{k\dist}, \frac{\vdec}{k\dist}\end{bmatrix}\trans
 \label{eqn:like_3d_mean} \\
  \cov_3 \,&=\, \begin{bmatrix*}[r]
    \cov_{\parallax\parallax} & \cov_{\parallax\propm} \\
    \cov_{\propm\parallax} & \cov_{\propm\propm} \\
  \end{bmatrix*} \
  \,=\,  \begin{bmatrix*}[r]
    \sigparallax^2                                           & \sigparallax\sigpmra\corparallaxpmra & \sigparallax\sigpmdec\corparallaxpmdec \\
    \sigparallax\sigpmra\corparallaxpmra       & \sigpmra^2                                          & \sigpmra\sigpmdec\corpmrapmdec \\
     \sigparallax\sigpmdec\corparallaxpmdec  & \sigpmra\sigpmdec\corpmrapmdec     & \sigpmdec^2  
   \end{bmatrix*}   \label{eqn:like_3d_cov}
\end{alignat}
\end{subequations}
where in equation~\ref{eqn:like_3d_mean} $m_\propm$ is the two-element vector with the velocities, and in equation~\ref{eqn:like_3d_cov} $\cov_{\parallax\parallax}$ is the scalar in the top-left, $\cov_{\propm\propm}$ is the $2\times 2$ matrix in the bottom-right, and the other two terms are the remaining two-element vectors.

I factorize the prior into a product of (1) a 2D Gaussian prior in
velocity conditional on the distance and direction \hp, $P(\vra,
\vdec \given \dist, \hp)$, which has mean $m_\tau(\dist, \hp)$ and covariance
$\cov_\tau(\dist, \hp)$ ($\tau$ for $\tau$ransverse), and (2) a 1D non-Gaussian prior over distance, $P(\dist \given \hp)$. In paper~VI the mean and covariance of the velocity prior were fit with a mock catalogue as a function of HEALpixel, and the distance prior was the generalized gamma distribution from paper~V.

Using Bayes' theorem and then factorizing, the unnormalized posterior is\footnote{For brevity I exclude symbols to represent the various parameters of the priors, but these can be considered represented by the direction $p$.}
\begin{subequations}\label{eqn:kinegeopost2}
\begin{alignat}{2}
  & P^*(\dist, \vra, \vdec \given \parallax, \pmra, \pmdec, \cov_3, \hp) \label{eqn:kinegeopost2_LHS} \\
  & \,=\ \underbrace{P(\parallax, \pmra, \pmdec \given \dist, \vra, \vdec, \cov_3)}_{\text{likelihood}} \ \underbrace{P(\vra, \vdec \given \dist, \hp) \ P(\dist \given \hp) }_{\text{prior}} \\
  & \,=\ P(\pmra, \pmdec \given \parallax, \dist, \vra, \vdec, \cov_2) P(\vra, \vdec \given \dist, \hp) \
  \underbrace{P(\parallax \given \dist, \cancel{\vra, \vdec}, \cov_{\parallax\parallax}) P(\dist \given \hp)}_{\text{geometric posterior}}  \label{eqn:kinegeopost2_fourterms}
\end{alignat}
\end{subequations}
where $\cov_2$ will be computed below, and the dependence on the velocity in the parallax likelihood (the penultimate term in the last line) drops out due to conditional independence.
The missing normalization constant is $1/P(\parallax, \propm)$. 
We see that the posterior has factorized into a product of the geometric posterior 
(recall that $\cov_{\parallax\parallax} \equiv \sigparallax$) and a pair of velocity/proper motion-dependent terms.
We now examine each of the terms in this posterior to identify how we can compute or sample this efficiently.

{\bf Step 1}: First term of equation~\ref{eqn:kinegeopost2_fourterms}, $P(\propm \given \parallax, \dist, v, \cov_2)$.\\
The first of the four terms on the right side of equation~\ref{eqn:kinegeopost2_fourterms} is the 2D likelihood in proper motion conditioned on the parallax, for given $v$ and $\dist$.
The conditional of a 3D Gaussian is a 2D Gaussian, and I denote this 2D proper motion likelihood with $\mathcal{N}_\propm(m_2, \cov_2)$. Thus
\begin{alignat}{2}
P(\propm \given \parallax, \dist, v, \cov_2) \,&=\, \mathcal{N}_\propm(m_2, \cov_2) \,=\, \frac{1}{\sqrt{\det(2\pi\cov_2)}}\,\exp(H)
\label{eqn:like_pm}
\end{alignat}
where the mean and covariance are\footnote{The new covariance is the inverse of the sum of the inverse covariances. The new mean is the inverse covariance weighted sum of the means, multiplied by the new covariance. See \url{https://statproofbook.github.io/P/mvn-cond.html}.}
\begin{subequations} \label{eqn:like_pm_mean_and_cov} 
\begin{alignat}{2}
  m_2 \,&=\, m_\propm + \underbrace{\cov_{\propm\parallax}\cov_{\parallax\parallax}\inv (\parallax - m_\parallax)}_{X_\propm} \label{eqn:like_pm_mean} \\
  \cov_2 \,&=\, \cov_{\propm\propm} - \cov_{\propm\parallax} \cov_{\parallax\parallax}\inv \cov_{\parallax\propm} \label{eqn:like_pm_cov}
\end{alignat}
\end{subequations}
 with terms computed from equation~\ref{eqn:like_3d}.
The term in the exponential of this Gaussian, $H$, can be rewritten (using equation~\ref{eqn:like_3d_mean} for $m_\propm$) as
\begin{alignat}{2}
  H \,&=\, -\frac{1}{2}\left[ \propm - \left(\frac{v}{k\dist} + X_\propm \right) \right] \cov_2\inv \left[\propm - \left(\frac{v}{k\dist} + X_\propm \right) \right]\trans \nonumber \\
  \,&=\, -\frac{1}{2}\,(v - k\dist[\mu - X_\propm])
  (k^2\dist^2\cov_2)\inv (v - k\dist[\mu - X_\propm])\trans \ .
  \label{eqn:like_pm_H}
\end{alignat}
Thus equation~\ref{eqn:like_pm} has the form of a Gaussian in $v$ (both the mean and covariance terms in equation~\ref{eqn:like_pm_H} are independent of $v$), which I'll call $\mathcal{N}_v$, but we need to be careful with the normalization constant. The normalization constant corresponding to equation~\ref{eqn:like_pm_H} is $1/\sqrt{\det(k^2 \dist^2\cov_2)}$, which differs from that in equation~\ref{eqn:like_pm} by a factor of $k^2 \dist^2$. Hence\footnote{Given a matrix $M$ of dimension $d$, for constant $a$ we have $\det(aM) = a^d \det(M)$. In our case $d=2$. Another way of looking at this is to consider transforming a PDF in \propm\ to one in $v$ using equation~\ref{eqn:transvel}
\begin{equation}
	\mathcal{N}_\propm \, d\propm 
	\,=\, \mathcal{N}_{\pmra \pmdec} \, d\pmra d\pmdec 
    \,=\, \mathcal{N}_{\vra \vdec} \, d\vra d\vdec 
    \,=\, \mathcal{N}_{\vra \vdec} k\dist d\pmra \, k\dist d\pmdec
    \,=\, k^2 \dist^2 \mathcal{N}_v d\propm \ . \nonumber
\end{equation}
} % end footnote
\begin{alignat}{2}
	P(\propm \given \parallax, \dist, v, \cov_2) 
	\,&=\, \mathcal{N}_\propm(m_2, \cov_2)
	\,=\,  k^2\dist^2 \, \mathcal{N}_v(k\dist[\mu - X_\propm], k\dist\cov_2) \ .
\end{alignat}
We have transformed from a Gaussian in \propm\ to a Gaussian in $v$, which is possible
because $m_\propm \propto v$ (equation~\ref{eqn:like_3d_mean}).
The factor of $k^2\dist^2$ arises from the change of variables (we see it preserves the units). Note that if the proper motion and parallax are uncorrelated (the likelihood factorizes), then $\cov_{\propm\parallax}=0$ (equation~\ref{eqn:like_3d_cov}) in equation~\ref{eqn:like_pm_mean}, hence $X_\propm=0$ and so this PDF becomes independent of the parallax.

{\bf Step 2}: First two terms of equation~\ref{eqn:kinegeopost2_fourterms}, $P(\propm \given \parallax, \dist, v, \cov_2) P(v \given \dist, \hp)$.\\
We now see this as a product of two Gaussians in $v$
\begin{alignat}{2}
  \frac{k^2\dist^2}{\sqrt{\det(2\pi k^2\dist^2\cov_2)\det(2\pi \cov_\tau)}}
  & \exp\left[ -\frac{1}{2} (v - k\dist[\mu - X_\propm]) (k^2\dist^2\cov_2)\inv (v - k\dist[\mu - X_\propm])\trans \right] \ \times \nonumber \\
  & \exp\left[ -\frac{1}{2} (v-m_\tau)\cov_\tau\inv (v-m_\tau)\trans \right] \ .
\end{alignat}
Recall that both $m_\tau$ and $\cov_\tau$ are functions of distance (and direction). 
By expanding and then completing the square, we can write
this\footnote{See \url{https://archive.org/details/imm3274/} (the
  matrix cookbook) section 8.1.8.} 
as a {\em non-normalized} Gaussian in $v$,
\begin{subequations}
  \begin{alignat}{2}
  Q(\dist, \parallax, \propm) & \, \mathcal{N}_v(m_v, \cov_v) \hspace*{1em}\textrm{where}\hspace*{1em} \\
  Q(\dist, \parallax, \propm) \,=&\,
  \frac{k^2\dist^2}{\sqrt{\det(2\pi[k^2\dist^2\cov_2 + \cov_\tau])}} \ \times
  \nonumber \\
  & \exp\left[ -\frac{1}{2}  (k\dist[\mu - X_\propm] - m_\tau) (k^2\dist^2\cov_2 + \cov_\tau)\inv  (k\dist[\mu - X_\propm] - m_\tau)\trans \right] \\
  \,=&\ k^2\dist^2 \, \mathcal{N}_{m_\tau}(k\dist[\mu - X_\propm], k^2\dist^2\cov_2 + \cov_\tau)
\end{alignat}
\end{subequations}
and
\begin{subequations}\label{eqn:post_vel}
\begin{alignat}{2}
  m_v   \,&=\,   ([k^2\dist^2\cov_2]\inv + \cov_\tau\inv)\inv \, ([k^2\dist^2\cov_2]\inv k\dist(\propm-X_\propm) + \cov_\tau\inv m_\tau) \label{eqn:post_vel_mean} \\
  \cov_v \,&=\,   ([k^2\dist^2\cov_2]\inv + \cov_\tau\inv)\inv \label{eqn:post_vel_cov} \ .
\end{alignat}
\end{subequations}
In addition to the explicit dependencies, $Q(\dist, \parallax, \propm)$ depends also on the covariances of the data and the mean and covariance of the velocity prior, but it does not depend on $v$. Note that $Q \mathcal{N}_v$ has units of $k^2\dist^2v^{-1} \times v^{-1} = \propm^{-1} v^{-1}$, which correctly retains the units of the original term.

{\bf Step 3}: The 3D joint distance--velocity posterior.\\
We can now write the complete kinegeometric posterior (expression~\ref{eqn:kinegeopost2_LHS}) as
\begin{subequations}\label{eqn:kinegeopost3}
\begin{alignat}{2} 
  P^*(\dist, v \given \parallax, \propm, \cov_3, \hp) \,&=\,
\underbrace{\mathcal{N}_v(m_v, \cov_v)}_{f(v, \dist)} \, 
\underbrace{Q(\dist, \parallax, \propm) \, P(\parallax \given \dist, \cov_{\parallax\parallax})
P(\dist \given \hp)}_{g(r)} \label{eqn:kinegeopost3_a} \\
  \,&=\, P^*(v \given \dist) \, P^*(\dist) \label{eqn:kinegeopost3_b}
\end{alignat}
\end{subequations}
which defines $f(v, \dist)$ and $g(r)$: that these are equal to the conditional velocity posterior
$P^*(v \given \dist)$ and the marginal distance posterior $P^*(\dist)$ 
respectively will be shown in the next two steps. Note that both of these functions depend on \parallax\ and \propm.

{\bf Step 4}: 1D marginal distance posterior.\\
Integrating equation~\ref{eqn:kinegeopost3} over $v$ gives the marginal posterior in distance. As the only $v$-dependent term is a normalized Gaussian in $v$, its integral is unity, leaving the three other terms:
\begin{subequations}
\begin{alignat}{2}
  P^*(\dist \given \parallax, \propm, \cov_3, \hp)
  \,&=\, \int P^*(\dist, v \given \parallax, \propm, \cov_3, \hp) \, dv \\
  \,&=\, Q(\dist, \parallax, \propm) \,
             \underbrace{P(\parallax \given \dist, \cov_{\parallax\parallax}) \, P(\dist \given \hp)}_{\text{geometric posterior}}
              \ \equiv \ g(r) \label{eqn:distance_posterior}\ .
\end{alignat}
\end{subequations}
This is a product of the geometric posterior with $Q$.
All terms depend on distance, so this is a non-trivial 1D function of distance that we can sample numerically.

\newpage
{\bf Step 5}: 2D marginal velocity posterior.\\
Integrating equation~\ref{eqn:kinegeopost3} over \dist\ gives the marginal posterior in velocity
\begin{subequations}
\begin{alignat}{2}
  P^*(v \given \parallax, \propm, \cov_3, \hp)
  \,&=\, \int P^*(\dist, v \given \parallax, \propm, \cov_3, \hp) \, d\dist \\
  \,&=\, \int f(v, \dist) \ g(\dist) \, d\dist \\
  \,&\simeq\, \frac{1}{N_g} \sum_{i=1}^{N_g} f(v, \dist_i) \hspace*{1em}\textrm{where}\hspace*{1em} \dist_i \in g(r) \ . \label{eqn:MC_integral}
\end{alignat}
\end{subequations}
This is a Monte Carlo integral: we draw $N_g$ distance samples from $g(r)$ and then take the average of $f$ computed at these.\footnote{In the formal definition, $g$ is normalized
\[
\frac{\int g(\dist)f(\dist) \, d\dist}{\int g(\dist) \, d\dist}  \,=\,  \langle f(\dist) \rangle \,\simeq\, \frac{1}{N}\sum_{i=1}^N f(\dist_i) \ ,
\]
but this is not required in practice because the set of samples is the same whether drawn from a normalized or an unnormalized distribution.}
As $f(v, \dist_i)$ is a Gaussian distribution in $v$ at the particular $\dist_i$, 
we see that the velocity posterior, marginalized over distance, is a sum of Gaussians. 
In practice we construct $P(v)$ by sampling $P(v \given \dist)$ ($=f$) at a set of $N_g$ distances $\{\dist_i\}$ that are representative of $P(\dist)$. Each distance drawn specifies a Gaussian $f(v, r_i)$ from which we then draw one velocity sample. This set of samples -- each is a vector $(\vra, \vdec)$ -- characterizes the 2D marginal velocity posterior from which we can also extract the two marginal 1D velocity posteriors. 
The procedure is summarized in Figure~\ref{fig:sampling_1D_sketch}.

% made by ./sampling_1D_sketch.svg
\begin{figure}[t]
	\begin{center}
		\includegraphics[width=0.55\textwidth, angle=0]{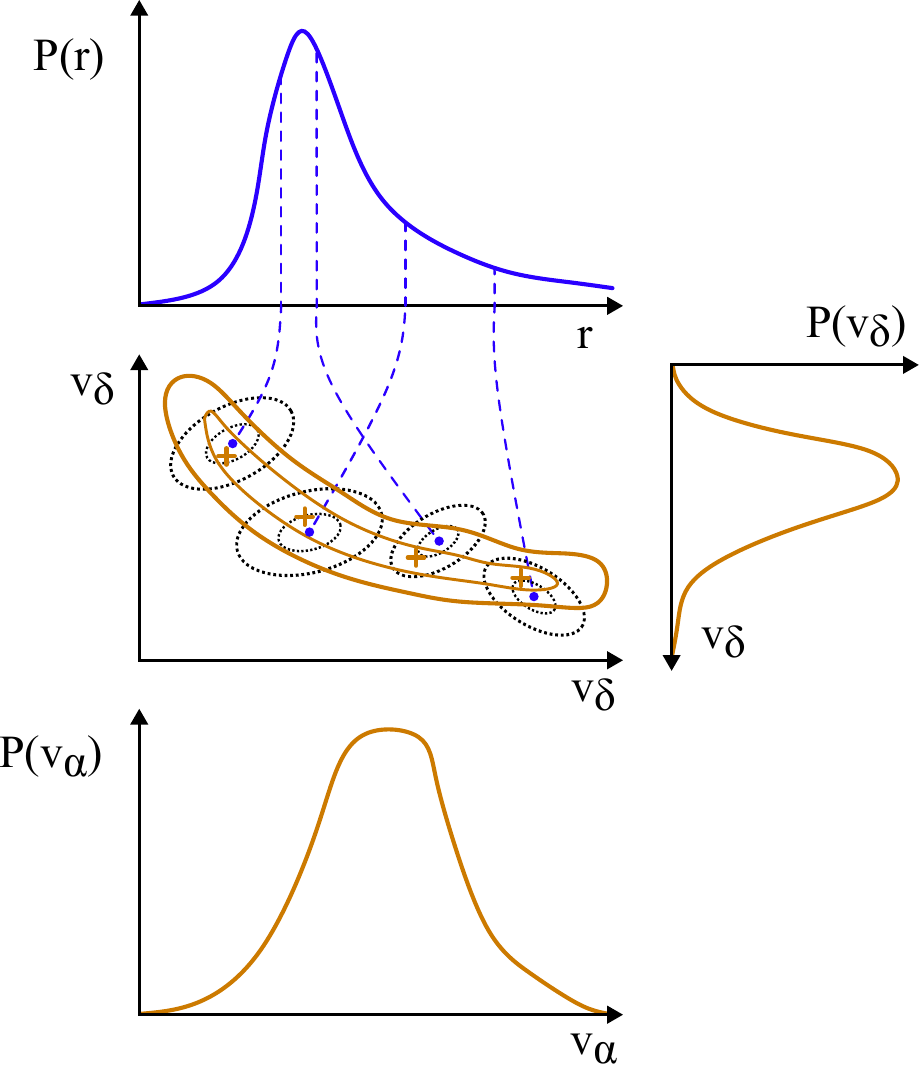}
		\caption{Sketch of the sampling procedure. Top: The 1D marginal distance posterior (equation~\ref{eqn:distance_posterior}), from which we draw $N_g$ distance samples $\{\dist_i\}$ (here $N_g=4$; typically we would use of order 1000). Centre left: Each $\dist_i$ (blue point) determines a conditional velocity posterior ($f$ in equation~\ref{eqn:kinegeopost3_a}), which is a 2D Gaussian posterior (black dashed contours). The marginal 2D velocity posterior is a sum of these (equation~\ref{eqn:MC_integral}). We draw one sample from each 2D Gaussian (orange cross) to estimate this marginal velocity posterior (orange contours). We can then view this as separate 1D marginal posteriors over \vra\ (bottom) and \vdec\ (centre right).
			\label{fig:sampling_1D_sketch} 
		}
\end{center}
\end{figure}

Finally, the conditional velocity posterior at a given distance \dist\ is
\begin{alignat}{2}
  P^*(v \given \dist) \,&=\, \frac{P^*(v,
    \dist)}{P^*(\dist)} \,=\, \mathcal{N}_v(m_v, \cov_v) \,\equiv\, f(v, \dist) \label{eqn:post_v_given_r}
\end{alignat}
which comes from dividing equation~\ref{eqn:kinegeopost3_a} by equation~\ref{eqn:distance_posterior}. This is
a 2D Gaussian with mean and covariance from equation~\ref{eqn:post_vel} evaluated at the
specified distance.

{\bf Step 6}: Posterior summaries.\\
From the previous steps we can produce a set of distance samples $N_g$ $\{\dist_i\}$ and a corresponding set of $N_g$ velocity samples  $\{v_i\}$. From these we may compute summary statistics, such as means and variances, 1D quantiles on the distance and velocities, as well as correlation coefficients. The distance and velocity are often strongly correlated because for a given proper motion, an increase in the distance can be compensated for by a decrease in the velocity. This correlation may make sampling the original 3D posterior difficult, but is irrelevant for the 1D+2D method derived here.

%%%%%%%%%%%%%%%%%%%%%%%%%%%%%%%%%%%%%%%%%%%%%%%%
\section{Summary of the procedure to sample the kinegeometric posterior}

\begin{enumerate}
	
	\item Sample the 1D marginal distance posterior $P^*(\dist \given \parallax, p)$
	($\equiv g(\dist))$
	in equation~\ref{eqn:distance_posterior} to provide a set of $N_g$ samples $\{\dist_i\}$.
	
	\item For each sample $\dist_i$, the {\em conditional} posterior
	velocity PDF $P^*(v \given \dist_i, \parallax, \propm, p)$
	($\equiv f(v, \dist_i)$) in equation~\ref{eqn:post_v_given_r} is a
	2D Gaussian in velocity with mean and covariance given by
	equation~\ref{eqn:post_vel} (using also equations~\ref{eqn:like_pm_mean_and_cov}
	and~\ref{eqn:like_3d}). The average of these $N_g$ Gaussians
	(equation~\ref{eqn:MC_integral}) gives the {\em marginal} 2D
	velocity posterior. We don't explicitly average the
	distributions. Instead, draw one sample of $(\vra, \vdec)$ from
	each of the $N_g$ functions $f(v , \dist_i)$. These samples represent the 2D marginal velocity
	posterior, from which we can compute summary statistics for
	velocity.
	
	\item The $N_g$ samples of distance and corresponding velocities can be used to characterize the 3D posterior, 
	for example by computing quantiles and correlation coefficients.
	
\end{enumerate}

%%%%%%%%%%%%%%%%%%%%%%%%%%%%%%%%%%%%%%%%%%%%%%%%
\section{Numerical demonstration}

An illustration of the application of this method to one simulated star is shown in Figure~\ref{fig:example_1d_sampling} with the orange lines and points.
The blue line and points shows the results from sampling the full 3D kinegeometric posterior. 
There is good agreement between the methods.

\begin{figure}
	\begin{center}
		\includegraphics[width=0.95\textwidth, angle=0]{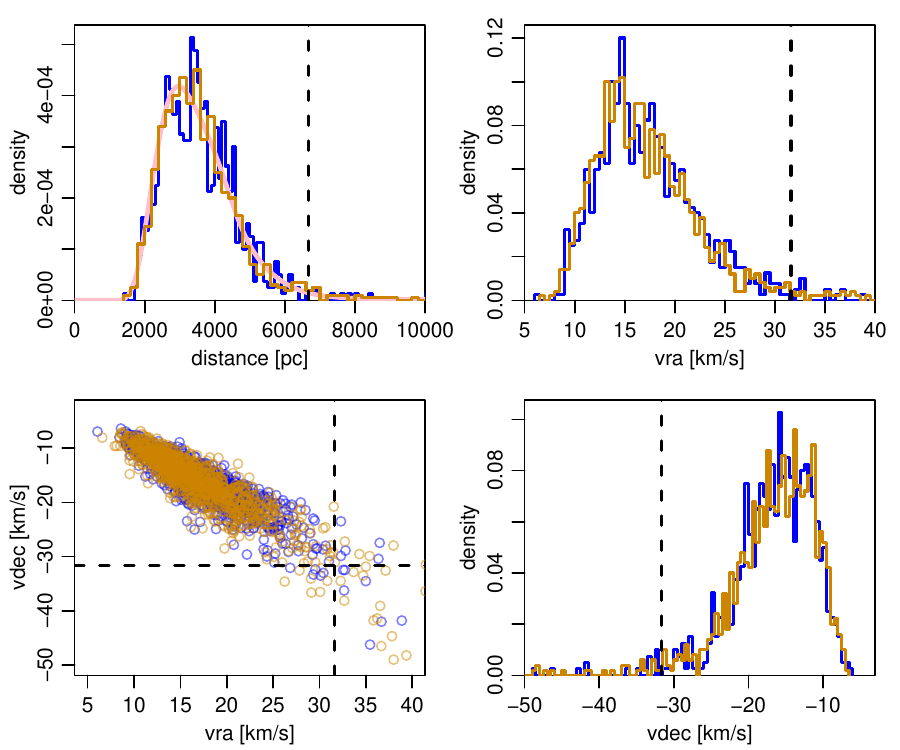}
		\caption{Kinegeometric posteriors for a star with $\parallax = 0.15 \pm 0.15$\,\mas, $\pmra = 1 \pm 0.1$\,\maspyr, and $\pmdec = -1 \pm 0.1$\,\maspyr, with small correlations. The blue histogram shows the results of the 3D sampling, the orange histogram the 1D+2D sampling method described in this paper (the smooth pink line in the top left panel shows this computed on a dense regular grid). The black dashed lines show the naive distance inference from $1/\parallax$ in the top-left panel, and the naive inference of velocities from $k\propm/\parallax$ in the other three panels. 
			\label{fig:example_1d_sampling} 
		}
	\end{center}
\end{figure}

For the 3D sampling in Figure~\ref{fig:example_1d_sampling} I used a parallel explorer MCMC algorithm, the emcee algorithm of Foreman-Mackey et al.\ (2013) with 20 walkers and a burn-in of 100 iterations, followed by 2000 samples that are thinned by a factor of ten.
To sample the 1D distance posterior in the 1D+2D method, I use the polynomial interpolation-based inversion of the cumulative distribution function (PINV) (Derflinger et al. 2010), as implemented in the Runuran library in R.\footnote{\url{https://doi.org/10.32614/CRAN.package.Runuran}}
This is not an MCMC, so does not suffer from problems of convergence or chain correlation. The interpolation is not perfect in principle, but is extremely good in practice.
PINV is relatively slow to setup, but extremely fast to sample; I use 2000 samples. The 2D Gaussian sampling to obtain the velocity posteriors is then done by the mvnfast library in R. \footnote{\url{https://doi.org/10.32614/CRAN.package.mvnfast}}.

I made a more extensive comparison of the 3D method with the 1D+2D method by applying both to estimate the posteriors of 500 stars in Gaia DR3, which have a broad range or parallaxes, proper motions, and covariances.
For both methods, I use the samples to compute the median, and the 16th and 84th percentiles of the distance and velocity posteriors as well as the correlation coefficients. For every source I then compute the ratio of each statistic between the two methods, and plot the distributions of these ratios in Figure~\ref{fig:sampling_comparison}.
 
 \begin{figure}[!t]
 	\begin{center}
 		\includegraphics[width=0.9\textwidth, angle=0]{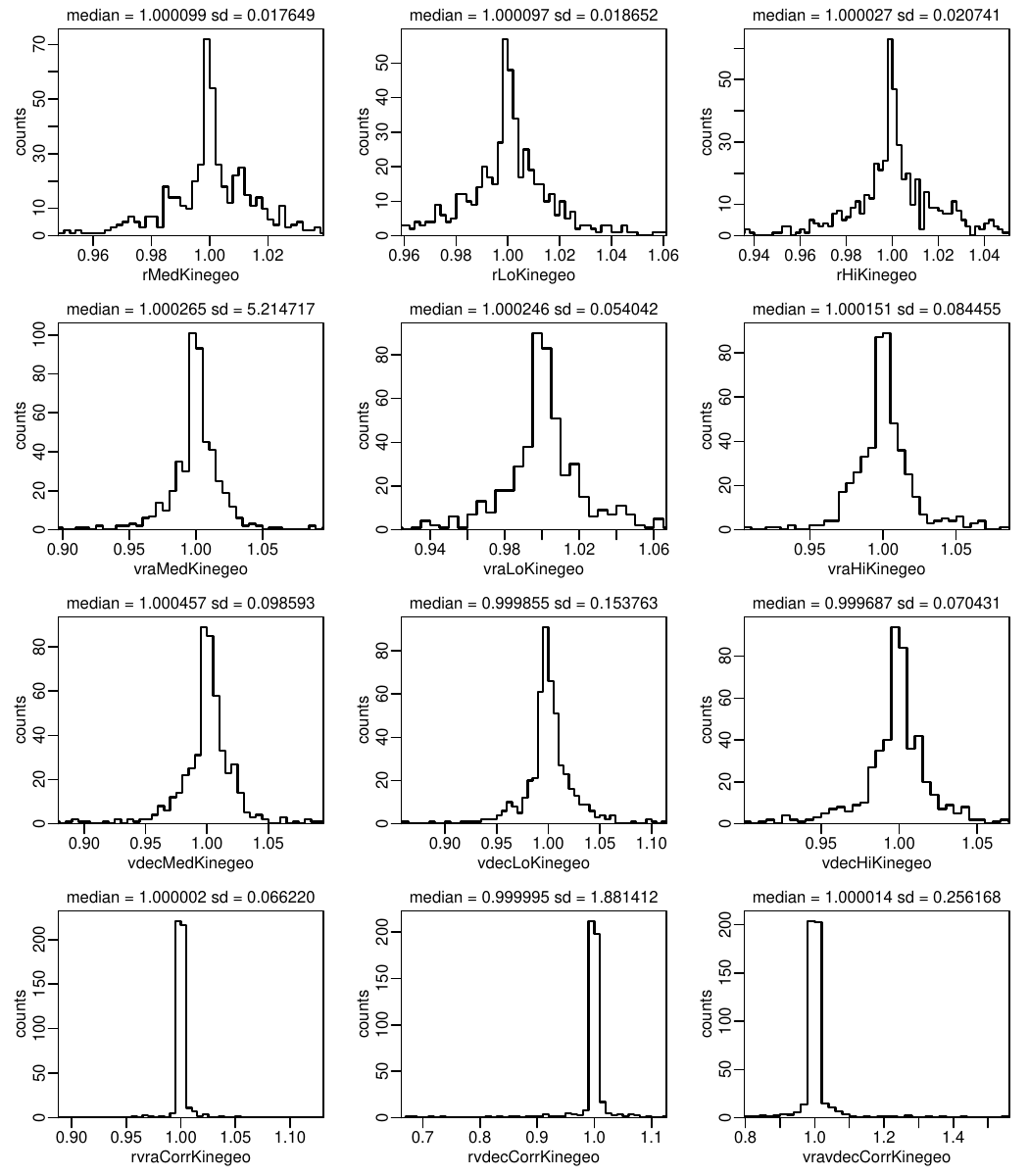}
 		\caption{Comparison of various summaries (one per panel) of the distance and velocity posteriors for 500 real stars computed using the 1D+2D and 3D methods. Each panel shows the distribution over the stars of the ratio of a posterior summary computed by the two methods (in the sense of 1D+2D divided by 3D). The top row shows the three distance quantiles (left to right: median, 16th percentile, 84th percentile) and the next two rows the quantiles for the two velocities. The bottom row shows the three correlation coefficients. The horizontal limits of each panel span the 1st and 99th percentiles of each distribution.
 			\label{fig:sampling_comparison} 
 		}
 	\end{center}
 \end{figure}
 
 If the methods agreed perfectly, the distributions of the ratios in Figure~\ref{fig:sampling_comparison} would all be delta functions at unity. They are quite close to this, as also quantified by the median and standard deviation written at the top of each panel.
Deviations from the ideal delta functions are presumably due to the finite number of samples used in both methods, plus imperfect convergence and residual correlation in the MCMC. 

 This demonstrates that the 1D+2D sampling method derived in this article is an accurate representation of the original 3D posterior. Furthermore, the 1D+2D method was 15 times faster to compute than the 3D MCMC method. This speed up comes primarily from the reduction of the dimensionality, not from the use of PINV in the 1D sampling. If we instead use emcee for this -- with 4 walkers, 100 burn-in, and 2000 samples thinned by a factor of ten -- it only takes 50\% longer than the PINV method (although in practice we may need more samples for good convergence). The speed up arises because we need fewer evaluations of a simpler posterior.

%%%%%%%%%%%%%%%%%%%%%%%%%%%%%%%%%%%%%%%%%%%%%%%%
\section{Conclusion}

To compute the 3D distance--velocity kinegeometric posterior we can draw $N_g$ samples from the marginal 1D non-Gaussian distance posterior, followed by the same number of samples from the conditional velocity posterior, which is a 2D Gaussian mixture model.
This approach is more accurate than directly sampling the 3D non-Gaussian joint distance--velocity posterior, and in practice requires far fewer posterior evaluations, making it faster.
MCMC could be used to sample the 1D distribution, but for these posteriors, 1D sampling can be done more efficiently at higher accuracy with other numerical methods.

%%%%%%%%%%%%%%%%%%%%%%%%%%%%%%%%%%%%%%%%%%%%%%%%
\section*{Acknowledgements}

No AI tools were used in this work, in particular not in the derivation or in the writing. I thank Rene Andrae for some useful discussions.

%%%%%%%%%%%%%%%%%%%%%%%%%%%%%%%%%%%%%%%%%%%%%%%%
\section*{References}

Bailer-Jones C.A.L., 2015. \href{http://adsabs.harvard.edu/abs/2015PASP..127..994B}{{\em Estimating distances from parallaxes: a tutorial}}. PASP 127, 994 [paper~I]

Bailer-Jones C.A.L., Rybizki J., Fouesneau M., Mantelet G., Andrae R., 2018.
\href{https://ui.adsabs.harvard.edu/abs/2018AJ....156...58B/abstract}{{\em Estimating distances from parallaxes. IV. Distances to 1.33 billion stars in Gaia data release 2}}. 
AJ 156, 58 [paper~IV]

Bailer-Jones C.A.L., Rybizki J., Fouesneau M., Demleitner M., Andrae R., 2021.
\href{https://www.mpia.de/homes/calj/gedr3_distances.html}{{\em Estimating distances from parallaxes. V. Geometric and photogeometric distances to 1.47 billion stars in Gaia Early Data Release 3}}. 
AJ 161, 147 [paper~V]

Bailer-Jones C.A.L., 2023.
\href{https://www.mpia.de/homes/calj/gedr3_distances_velocities.html}{{\em Estimating distances from parallaxes.
VI. A method for inferring distances and transverse velocities from parallaxes and proper motions demonstrated on Gaia Data Release 3}}.
AJ 166, 269 [paper~VI]

Derflinger G., Hörmann W., Leydold J., 2010.
\href{https://doi.org/10.1145/1842722.1842723}{{\em Random variate generation by numerical inversion when only the density is known}}.
ACM Transactions on Modeling and Computer Simulation (TOMACS), 20(4), 18

Foreman-Mackey D., Hogg D.W., Lang D., Goodman J., 2013. \href{https://ui.adsabs.harvard.edu/abs/2013PASP..125..306F/abstract}{{\em emcee: The MCMC Hammer}}.
PASP 125, 925

Gaia Collaboration, 2018. \href{https://ui.adsabs.harvard.edu/abs/2018A%26A...616A...1G/abstract}{{\em Gaia Data Release 2. Summary of the contents and survey properties}}.
A\&A 616, A1

Gaia Collaboration, 2021. \href{https://ui.adsabs.harvard.edu/abs/2021A%26A...649A...1G/abstract}{{\em Gaia Early Data Release 3. Summary of the contents and survey properties}}.
  A\&A 649, A1

Lindegren L., Hernández J., Bombrun A., Klioner S., Bastian U., et al.\ 2018
\href{https://ui.adsabs.harvard.edu/abs/2018A%26A...616A...2L/abstract}{{\em Gaia Data Release 2. The astrometric solution}}.
A\&A 616, A2

 Luri X., Brown  A.G.A., Sarro L. M., Arenou F., Bailer-Jones C. A. L., et al.\ 2018. 
 \href{https://ui.adsabs.harvard.edu/abs/2018A%26A...616A...9L/abstract}{{\em Gaia Data Release 2. Using Gaia parallaxes}}.
 A\&A 616, A9

 Queiroz A.B.A., Anders F., Santiago B.X., Chiappini C., Steinmetz M., et al., 2018.
  \href{https://ui.adsabs.harvard.edu/abs/2018MNRAS.476.2556Q/abstract}{{\em StarHorse: a Bayesian tool for determining stellar masses, ages, distances, and extinctions for field stars}}.
 MNRAS 476, 2556

\end{document}